\documentclass[12pt,aps,prd,preprint,tightenlines,
   showpacs,nofootinbib]{revtex4-1}
\newcommand{\PRE}[1]{{#1}} 

\usepackage{hyperref}      
\usepackage{bm}
\usepackage{epsfig}

\def\beq{\begin{eqnarray}}
\def\eeq{\end{eqnarray}}
\def\bea{\begin{eqnarray}}
\def\eea{\end{eqnarray}}

\newcommand{\mweak}{m_W}
\newcommand{\mgut}{m_{\text{GUT}}}

\newcommand{\ifb}{\text{fb}^{-1}}

\newcommand{\gev}{\text{GeV}}
\newcommand{\tev}{\text{TeV}}

\newcommand{\eqref}[1]{Eq.~(\ref{#1})}

\newcommand{\Eqref}[1]{Equation~(\ref{#1})}
\newcommand{\secref}[1]{Sec.~\ref{sec:#1}}

\newcommand{\figref}[1]{Fig.~\ref{fig:#1}}
\newcommand{\figsref}[2]{Figs.~\ref{fig:#1} and \ref{fig:#2}}
\newcommand{\Figref}[1]{Figure~\ref{fig:#1}}

\newcommand{\mgaugino}{M_{1/2}}

\newcommand{\mhu}{m_{H_u}}
\newcommand{\mhd}{m_{H_d}}
\newcommand{\mtl}{m_{Q_3}}
\newcommand{\mtr}{m_{U_3}}
\newcommand{\at}{A_t}

\newcommand{\ab}{A_b}

\newcommand{\mmix}{\mathcal{A}_{\tilde{t}}}

\begin{document}

\preprint{UCI-TR-2012-12}

\title{ \PRE{\vspace*{1.5in}} A Natural 125 GeV Higgs Boson in the
  MSSM from Focus Point Supersymmetry with $A$-Terms
  \PRE{\vspace*{0.3in}} }

\author{Jonathan L.~Feng}
\affiliation{Department of Physics and Astronomy, University of
California, Irvine, CA 92697, USA
\PRE{\vspace*{.4in}}
}

\author{David Sanford\PRE{\vspace*{.1in}}}
\affiliation{Department of Physics and Astronomy, University of
California, Irvine, CA 92697, USA
\PRE{\vspace*{.4in}}
}

\begin{abstract}
\PRE{\vspace*{.3in}} We show that a 125 GeV Higgs boson and
percent-level fine-tuning are simultaneously attainable in the MSSM,
with no additional fields and supersymmetry breaking generated at the
GUT scale. The Higgs mass is raised by large radiative contributions
from top squarks with significant left-right mixing, and naturalness
is preserved by the focus point mechanism with large $A$-terms, which
suppresses large log-enhanced sensitivities to variations in the
fundamental parameters.  The focus point mechanism is independent of
almost all supersymmetry-breaking parameters, but is predictive in the
top sector, requiring the GUT-scale relation $\mhu^2 : \mtr^2 : \mtl^2
: \at^2 = 1 : 1+x - 3y : 1-x : 9y$, where $x$ and $y$ are constants.
We derive this condition analytically and then investigate three
representative models through detailed numerical analysis.  The models
generically predict heavy superpartners, but dark matter searches in
the case of non-unified gaugino masses are promising, as are searches
for top squarks and gluinos with top and bottom-rich cascade decays at
the LHC.  This framework may be viewed as a simple update to
mSUGRA/CMSSM to accommodate both naturalness and current Higgs boson
constraints, and provides an ideal framework for presenting new
results from LHC searches.

\end{abstract}

\pacs{12.60.Jv, 14.80.Da, 95.35.+d}

\maketitle

\section{Introduction}  

For three decades, weak-scale supersymmetry (SUSY) has been strongly
motivated by three main promises: a natural solution to the gauge
hierarchy problem, an excellent dark matter candidate, and force
unification.  Recent results from the LHC with center-of-mass energy
$\sqrt{s} = 7~\tev$ and integrated luminosity $\sim 5~\ifb$ have begun
to challenge this paradigm. The challenge arises from two sources:
first, null results from superpartner searches require the gluino and
some squarks to have masses around 1 TeV or above in conventional SUSY
scenarios~\cite{ATLAS-CONF-2012-033,CMS-PAS-SUS-12-002}, and second,
the ATLAS and CMS experiments have excluded much of the Higgs boson
mass range and, at the same time, have reported excesses consistent
with a standard model (SM)-like Higgs boson at masses of
$126~\gev$~\cite{ATLAS:2012ae} and
$124~\gev$~\cite{Chatrchyan:2012tx}, respectively.

The null results from superpartner searches are the most direct
constraints, but they are not especially problematic for weak-scale
SUSY.  Thermal relic neutralinos and gauge coupling unification
provide only weak upper bounds on the masses of superpartners, and are
completely consistent with heavy scalars far above the TeV
scale~\cite{Feng:2000gh,Feng:2000bp}.  With regard to naturalness,
TeV-scale superpartners generically require percent-level fine-tuning
of the weak scale.  Although naturalness is a notoriously subjective
and brittle concept, we consider such fine-tuning acceptable.  Less
subjective is the fact that, for many superpartners, there are good
reasons to expect them to be far above current LHC bounds.  For
example, for the first two generations of squarks, $10~\tev$ masses
are
natural~\cite{Drees:1985jx,Dine:1993np,Dimopoulos:1995mi,Pomarol:1995xc},
flavor constraints generically require masses far above the TeV scale,
and even in models that automatically conserve flavor, electric dipole
moments typically require superpartner masses well above 1
TeV~\cite{Gabbiani:1996hi}.  In light of these longstanding facts, the
fact that superpartners have not yet been discovered at the LHC should
not be a great surprise.

The Higgs boson results, although still tentative, are more troubling.
Although a 125 GeV Higgs boson is, broadly speaking, in the range
expected for SUSY, it typically requires large radiative corrections
from top squarks, and the required stop properties generically induce
large fine-tuning.  For example, in the case of the minimal
supersymmetric standard model (MSSM), with SUSY-breaking mediated at
the grand unification theory (GUT) scale and negligible left-right
stop mixing, the required stop masses are around 10 TeV, typically
corresponding to a fine-tuning of roughly 1 part in $10^4$.  With
maximal mixing, the stops may be lighter, and the fine-tuning may be
reduced to roughly 1 part in 1000~\cite{Hall:2011aa}.  Such results
are extremely sensitive to the actual value of the Higgs boson mass,
as well as to still significant uncertainties in the theoretical
calculation of the Higgs boson mass in SUSY, and their interpretation
is again subject to individual taste.  The tension has, however,
motivated numerous re-examinations of naturalness (see, for example,
Refs.~\cite{Ellis:2012aa,Baer:2012uy,Ghilencea:2012gz,Baer:2012uy2}),
as well as many reconsiderations of extensions of the MSSM, with all
their attendant difficulties.

It is important to note, however, that the apparent conflict between
the 125 Higgs boson and naturalness assumes that the soft SUSY
parameters are uncorrelated. SUSY theories with uncorrelated soft
parameters are, however, excluded --- a wealth of experimental data
implies that if weak-scale SUSY exists, there must be structure behind
the soft parameters.  The possibility that correlations between soft
parameters reconcile naturalness with heavy superpartners is
formalized in the framework of focus point (FP)
SUSY~\cite{Feng:1999mn,Feng:1999zg}.  In FP SUSY, correlations reduce
the sensitivity of the weak scale to variations in the fundamental
parameters, even if they are large and above the TeV scale. This
insensitivity may be understood in a number of equivalent ways.
Graphically, in FP SUSY, the insensitivity may be understood as a
property of renormalization group (RG) trajectories, which focus to a
fixed value at the weak scale independent of their value in the
ultraviolet.  Alternatively, FP SUSY may be understood as suppressing
the large log-enhanced sensitivity to GUT-scale parameters, leaving
only the ``irreducible'' quadratic sensitivity, and thereby reducing
fine-tuning by factors of roughly $\ln(\mgut/m_{\tilde{t}}) \sim
30$. From any view, identifying naturalness with insensitivity, all
natural theories with high-scale mediation and multi-TeV top squarks
are FP models, and current results from Higgs boson searches at the
LHC provide a strong motivation for FP SUSY.

Early analyses of the FP
mechanism~\cite{Feng:1999hg,Feng:1999mn,Feng:1999zg} considered the
MSSM with multi-TeV scalar masses, but small $A$-terms.  This
framework was reviewed recently in light of the LHC Higgs
results~\cite{Feng:2011aa}.  Depending on Higgs mass uncertainties,
regions of parameter space that are fine-tuned to as little as 1 part
in 500 are consistent with the Higgs search results, as we will review
below.  Although some scalars are very heavy, these models have some
superpartners well within the reach of the LHC, notably the gluino,
and excellent WIMP dark matter candidates. The collider and
cosmological implications have been explored in many studies (see, for
example, Refs.~\cite{Feng:2000gh,Feng:2000zu,Chattopadhyay:2000qa,%
  Mercadante:2005vx,Feng:2005hw,DeSanctis:2007td,Das:2007jn,%
  Kadala:2008uy,Feng:2011aa}), and have implications not only for FP
SUSY, but also for all other models with heavy scalars.  FP SUSY has
also been investigated in many other related contexts, including
hyperbolic branch SUSY~\cite{Chan:1997bi,Akula:2011jx}, gauge-mediated
supersymmetry breaking models~\cite{Agashe:1999ct}, mirage
mediation~\cite{Kitano:2005wc}, models with large gaugino
masses~\cite{Abe:2007kf,Horton:2009ed,Younkin:2012ui}, and the MSSM
with right-handed neutrinos~\cite{Asano:2011kj}.

In this work, we consider FP models in which both scalar masses and
$A$-terms are multi-TeV.  The large $A$-terms allow for significant
stop mixing, but of course, requires a re-analysis of the FP
mechanism, since the $A$-terms can no longer be neglected in the RG
evolution.  We begin by reviewing the standard definition of
fine-tuning in \secref{finetune}, and comment on its implications and
some alternative definitions used in the literature.  We then present
an analytic derivation of focus points with large $A$-terms in
\secref{focuspoints}, deriving the necessary relationships between the
soft SUSY-breaking parameters.  With these analytic results as a
guide, we perform fully detailed numerical analyses in
\secref{numerical}.  We consider three representative models in
detail, and show that a 125 GeV Higgs mass may be obtained with only
percent-level fine-tuning in the MSSM, without additional field
content and without invoking a low mediation scale.  We present
implications for collider and dark matter searches in
\secref{experimental} and summarize our results in
\secref{conclusions}.

\section{Fine-Tuning in the MSSM} 
\label{sec:finetune}

For $\tan\beta \agt 5$, the tree-level condition for electroweak
symmetry breaking in the MSSM is
\begin{equation}
m_Z^2 \approx - 2 \mu^2 - 2 \mhu^2 (\mweak) \ ,
\label{mz2}
\end{equation}
where $\mhu^2 (\mweak)$ is the up-type Higgs mass parameter at the
weak scale $\mweak \sim 100~\gev - 1~\tev$, and $\mu$ is the Higgsino
mass parameter.  Natural SUSY theories generally fall into two
classes: conventional theories in which the fundamental parameters
determining $\mhu^2 (\mweak)$ have values at or below the TeV-scale
throughout their RG evolution, and FP theories, in which natural
values of $\mhu^2 (\mweak)$ are dynamically generated and insensitive
to the values of the GUT-scale parameters, even if these GUT-scale
parameters are significantly above the TeV scale.  In this study, we
restrict ourselves to the MSSM, that is, the supersymmetric model with
minimal field content, with soft SUSY-breaking scalar and gaugino
masses generated at the GUT scale $\mgut \simeq 2.4 \times
10^{16}~\gev$.

To evaluate naturalness, we define the sensitivity coefficients
\begin{equation}
c_a \equiv \left| \frac{\partial \ln m_Z^2}{\partial \ln a^2}\right| \ ,
\label{ca}
\end{equation}
where $a^2$ is one of the input GUT-scale parameters, including
$m_0^2$, $\mgaugino^2$, $\mu_0^2$, and $m_3^2$, the $H^0_u H^0_d$ mass
parameter.\footnote{We choose the GUT-scale parameter to be $m_0^2$,
  not $m_0$, because we consider $m_0^2$ to be more fundamental (it
  may be negative, for example~\cite{Feng:2005ba}), and because we
  consider it more reasonable to compare squared masses against one
  another, given \eqref{mz2}.  For this reason, we also choose $m_Z^2$
  instead of $m_Z$ in the numerator of \eqref{ca}, and for uniformity,
  choose all $a^2$ to be mass dimension two.  As a result, our
  definition differs from the original definition of $c_a = \partial
  \ln m_Z^2 / \partial \ln a$~\cite{Ellis:1986yg,Barbieri:1987fn} by a
  factor of 2.  Such factors are clearly unimportant in judging
  whether a scenario is natural or not, given the subjective nature of
  the definition, but are important to keep in mind when comparing
  numerical results.}  The overall fine-tuning of a model is defined
as
\begin{equation}
c \equiv \text{max}\{c_a \} \ .
\end{equation}
In the models we will consider, either $c_{m_0}$ or $c_{\mgaugino}$
determines $c$ in the interesting regions of parameter space.

Note that it is quite possible to have large values of the GUT-scale
parameters, but to arrange for $\mhu^2 (\mweak) \sim m_Z^2$ by
suitably fine-tuning values for these GUT parameters.  In such
scenarios, \eqref{mz2} may appear natural, and $c_{\mu_0} \propto
\mu^2 / m_Z^2$ will be low, but this should not obscure the fact that
the model has been fine-tuned to get low $\mhu^2$ and the weak scale
is nevertheless unnaturally sensitive to variations in the GUT-scale
parameters.  Here we require not just low $c_{\mu_0}$, but low $c$,
and the FP models discussed below will be natural according to this
stricter definition.

In the sections below, we will consider models with heavy scalars.
Below the scalar masses, the Higgs mass receives quadratic
contributions of the form $(6/ 8\pi^2) y_t^2 m_{\tilde{t}}^2$, where
$y_t$ is the top Yukawa coupling.  This contribution is {\em not} the
usual source of the fine-tuning problem --- it is one-loop suppressed,
and if this were all there were, even values as large as
$m_{\tilde{t}} \sim 3~\tev$ would only be percent-level fine-tuned.
The dominant source of fine-tuning, and the apparent conflict between
naturalness and the 125 GeV Higgs boson, is the large log-enhanced
contributions.  These result from RG evolution from the GUT scale and
are of the order of $\sim (6 / 8\pi^2) y_t^2 m_{\tilde{t}}^2
\ln(\mgut/m_{\tilde{t}})$.  The large logarithm roughly cancels the
loop-suppressed prefactor, leading to the conventional wisdom that
multi-TeV top squarks imply sub-percent-level fine-tuning.  To reduce
the dominant source of fine-tuning, then, one may consider the RG
equations (RGEs) and look for correlations that reduce the sensitivity
of the weak scale to variations in the GUT-scale parameters.  This is
the possibility formalized in the FP framework, to which we turn in
the next section. Of course, the ``irreducible'' quadratic
contribution will remain, and will be accounted for when fine-tuning
is evaluated through the full numerical analysis of
\secref{numerical}, in which two-loop RGEs and one-loop threshold
corrections are used and superpartners are integrated out at the
appropriate mass scale.

\section{Focus Points for Large Scalar Masses and $A$-Terms} 
\label{sec:focuspoints}

We begin in this section with a simple analytic discussion to extract
the desired FP behavior.  In \secref{numerical}, we verify the
validity of the results derived here through a full numerical
analysis.

\subsection{Renormalization Group Equations}

The 1-loop RGEs for SUSY parameters have the schematic form
\begin{eqnarray}
\frac{dg}{d \ln Q} &\sim& - g^3 \\
\frac{dy}{d \ln Q} &\sim& - g^2 y + y^3 \\
\frac{dM}{d \ln Q} &\sim& - g^2 M \\
\frac{dA}{d \ln Q} &\sim&  g^2 M + y^2 A \\
\frac{dm^2}{d \ln Q} &\sim& - g^2 M^2 + y^2 A^2 + y^2 m^2 \ ,
\end{eqnarray}
where positive numerical coefficients have been omitted, and $g$, $y$,
$M$, $A$, and $m$ are generic symbols for gauge couplings, Yukawa
couplings, gaugino masses, trilinear scalar couplings, and scalar
masses, respectively.

Because the scalar masses and $A$ parameters do not enter the gaugino
mass RGEs, it is self-consistent to assume $m^2, A^2 \gg M^2$ through
the RG evolution.  With this assumption, and further neglecting all
Yukawa couplings other than $y_t$,\footnote{This assumption is valid
  for low and moderate values of $\tan\beta$.  Extending this
  discussion to high $\tan\beta$ is possible, but requires unified
  $A$-terms for third generation sfermions ($\tilde{t}$, $\tilde{b}$,
  $\tilde{\tau}$) for the RGEs to be linear.  Doing so further
  requires the inclusion of a right-handed neutrino supermultiplet for
  consistent RGEs.} the RGEs reduce to
\begin{equation}
\frac{d}{d \ln Q} \left[ \begin{array}{c}
\mhu^2 \\ \mtr^2 \\ \mtl^2 \\ \at^2
\end{array} \right]
= \frac{y_t^2}{8 \pi} \left[ \begin{array}{cccc}
3 & 3 & 3 & 3 \\
2 & 2 & 2 & 2 \\
1 & 1 & 1 & 1 \\
0 & 0 & 0 & 12
\end{array} \right]
\left[ \begin{array}{c}
\mhu^2 \\ \mtr^2 \\ \mtl^2 \\ \at^2
\end{array} \right] \ .
\label{eq:fullrge}
\end{equation}
Here, we have neglected not only the gaugino masses, but also terms
proportional to $g_1^2 S$, where
\begin{equation}
\label{eq:S}
S = \mhu^2 - \mhd^2 + \text{tr}[\bm{m_Q^2} -
  \bm{m_L^2} - 2 \bm{m_U^2} + \bm{m_D^2} +
  \bm{m_E^2}] \ .
\end{equation}
These contributions will be discussed further in \secref{numerical}.

\Eqref{eq:fullrge} may be solved in terms of the eigenvalues and
eigenvectors of the $4 \times 4$ matrix of numerical coefficients.
The solution is
\begin{eqnarray}
\left[ \begin{array}{c} 
\mhu^2 (Q) \\ \mtr^2 (Q) \\ \mtl^2 (Q) \\ \at^2 (Q)
\end{array} \right] & = &
\kappa_{12} \left[ \begin{array}{c}
3 \\ 2 \\ 1 \\ 6
\end{array} \right] e^{12I(Q)} + 
\kappa_{6} \left[ \begin{array}{c}
3 \\ 2 \\ 1 \\ 0
\end{array} \right] e^{6I(Q)} + \kappa_{0} \left[ \begin{array}{c}
1 \\ 0 \\ -1 \\ 0
\end{array} \right] + 
\kappa_{0}' \left[ \begin{array}{c}
0 \\ 1 \\ -1 \\ 0
\end{array} \right] ,
\label{eq:solution}
\end{eqnarray}
where 
\begin{equation}
I(Q) = \int_{\ln Q_0}^{\ln Q} \frac{y_t^2 (Q')}{8 \pi^2} d \ln Q' 
\end{equation}
is a renormalization factor related to the running of the top Yukawa
coupling from the mass generation scale $Q_0$ to the scale $Q$.  It
takes the value $e^{6I(\mweak)} \simeq \frac{1}{3}$ for
renormalization between the GUT and weak scales~\cite{Feng:1999zg}.

To consider the possibility that a large value of $\mhu^2$ at the GUT
scale evolves to a much smaller value at the weak scale, we set
\begin{eqnarray}
m_0^2 &=& \mhu^2 (\mgut) = 3 \kappa_{12} + 3 \kappa + 
\kappa_0 \label{eq:mhugut} \\
0 &=& \mhu^2(\mweak) = 3 \kappa_{12} e^{12 I(\mweak)} + 
3 \kappa e^{6 I(\mweak)} + \kappa_0 \ . \label{eq:mhuzero}
\end{eqnarray}
With these conditions, and using the above approximation for
$e^{6I(\mweak)}$, the parameters evolve from the GUT scale to the
weak scale through
\begin{equation}
\left[ \begin{array}{c} 
\mhu^2 (\mgut) \\ \mtr^2 (\mgut) \\ \mtl^2 (\mgut) \\ \at^2 (\mgut)
\end{array} \right] =
m_0^2  \left[ \begin{array}{c} 
1 \\ 1+x - 3y \\ 1-x \\ 9y
\end{array} \right] 
\to
\left[ \begin{array}{c} 
\mhu^2 (\mweak) \\ \mtr^2 (\mweak) \\ \mtl^2 (\mweak) \\ \at^2 (\mweak)
\end{array} \right] =
m_0^2  \left[ \begin{array}{c} 
0 \\ \frac{1}{3} + x - 3y \\ \frac{2}{3} - x \\ y
\end{array} \right] .
\end{equation}
There is, of course, freedom in choosing the parameterization.  We
choose $x$ to parameterize the splitting between $\mtl^2 (\mgut)$ and
$\mtr^2 (\mgut)$, and $y$ to be directly related to $\at (\mgut)$.

\subsection{Model Parameter Space}

The parameter space therefore consists of an overall scale $m_0$, and
the two numbers $x$ and $y$.  The physically viable region of $(x,y)$
parameter space is determined by two constraints.  First, we require
$\at^2 (\mweak) \ge 0$, and so $y \ge 0$. Second, we require that both
stops are not tachyonic at the weak scale.  In the limit of large
scalar masses, $\mtr^2 (\mweak), \mtl^2 (\mweak) \gg m_t \at (\mweak)$
and one-loop corrections are subdominant, so to a good approximation,
the physical stop masses are $\mtr (\mweak)$ and $\mtl (\mweak)$.
Using this approximation, we find that the viable region is
\begin{eqnarray}
0 \le \ & y & \ \le \frac{1}{3} \\
-\frac{1}{3} + 3 y \le \ & x & \ \le \frac{2}{3} \ .
\label{eq:range}
\end{eqnarray}
This parameter space is shown in \figref{xy}.

\begin{figure}[tb]
\includegraphics[width=0.47\columnwidth]{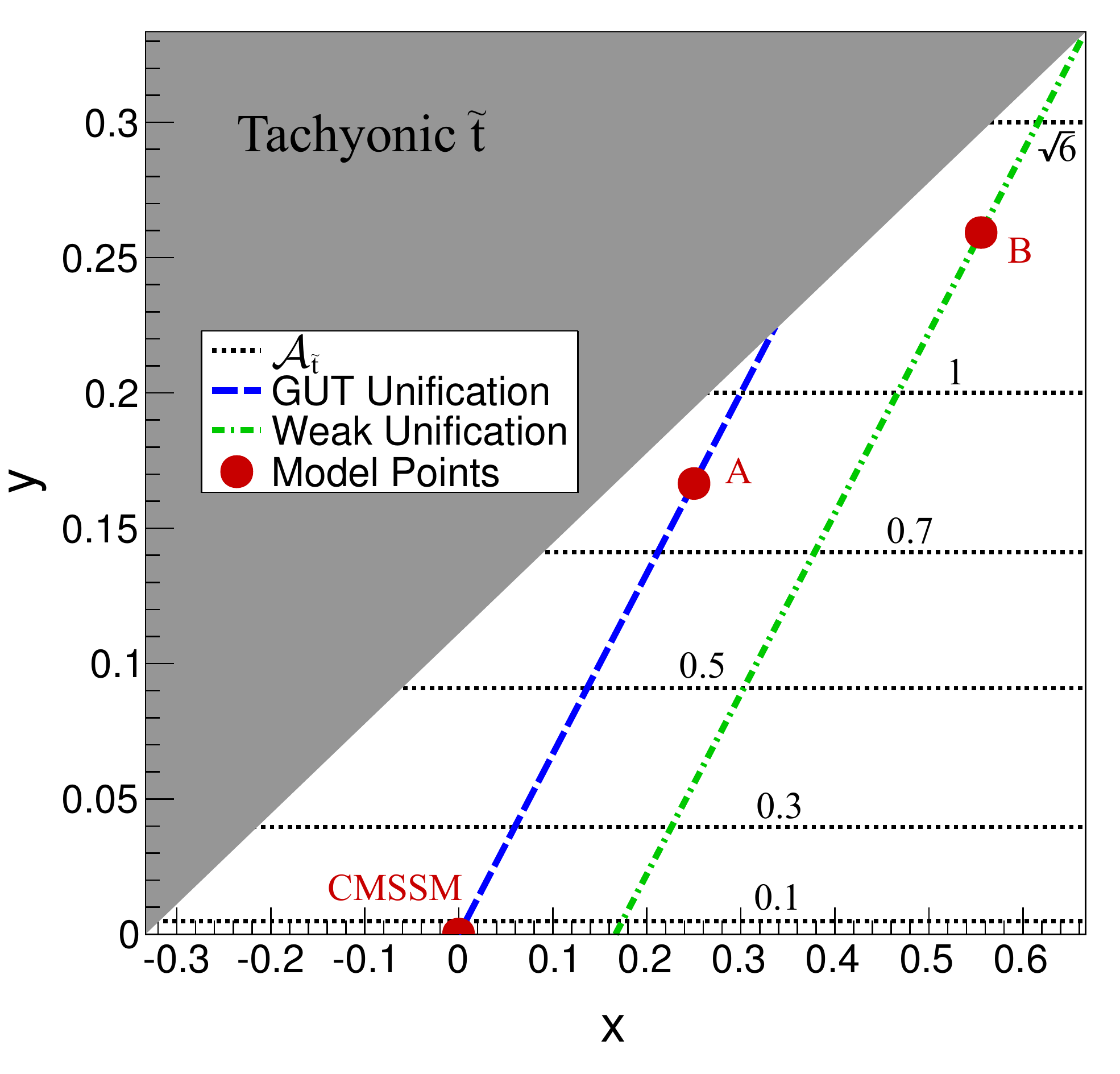}
\hfil
\includegraphics[width=0.51\columnwidth]{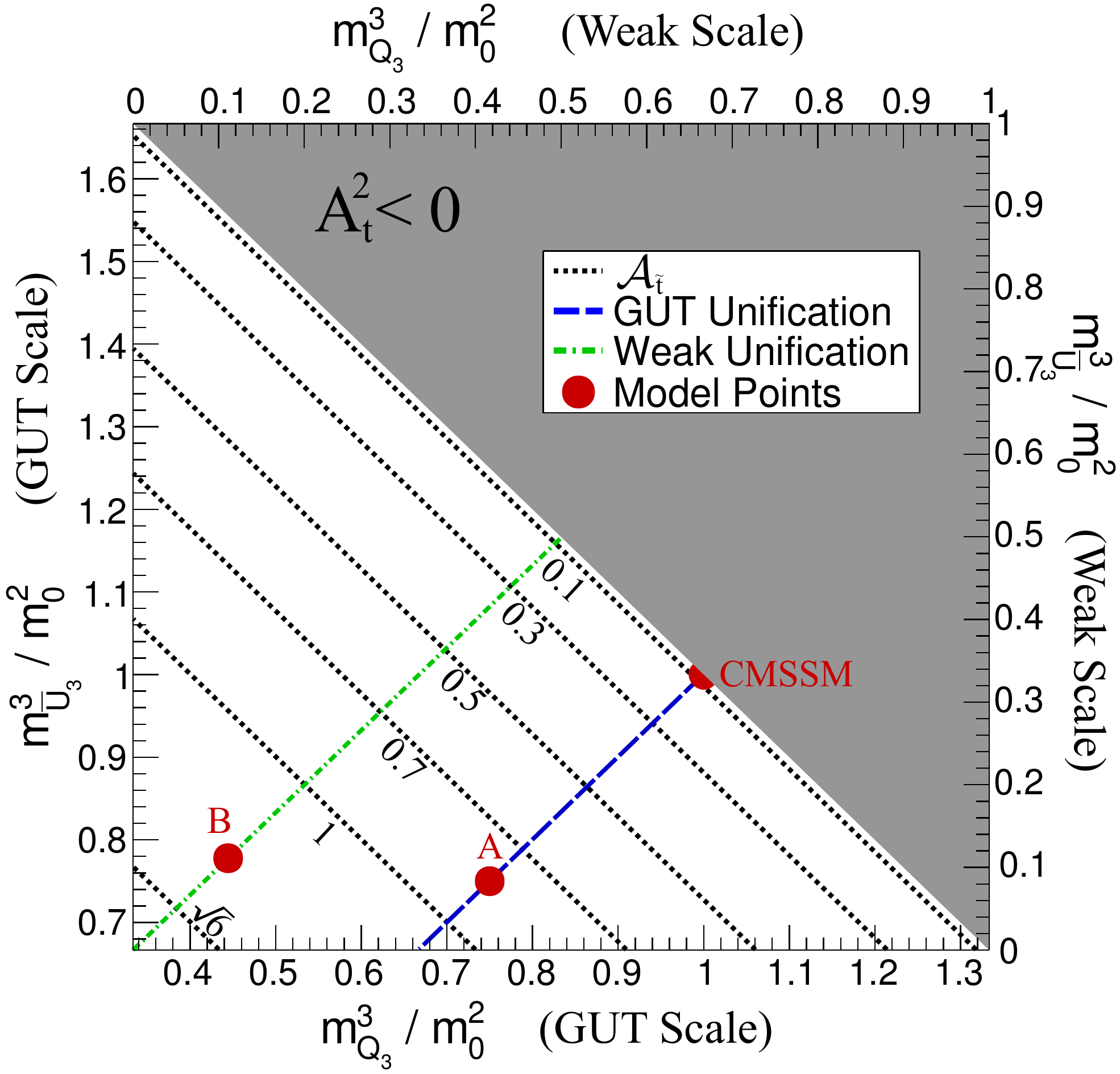}
\vspace*{-.1in}
\caption{\label{fig:xy} The model parameter space in the $(x,y)$ plane
  (left) and in the $(\mtl^2(\mgut), \mtr^2(\mgut))$ [or
    alternatively, $(\mtl^2(\mweak), \mtr^2(\mweak))$] plane (right).
  The blue dashed and green dot-dashed lines correspond to the special
  cases where $\mtl^2 = \mtr^2$ at the GUT and weak scales,
  respectively, and red dots correspond to the three models $(x,y) =
  (0,0)$ (mSUGRA/CMSSM), $(\frac{1}{4}, \frac{1}{6})$, and
  $(\frac{5}{9}, \frac{7}{27})$ examined in detail.  Black dotted
  lines of constant stop mixing parameter $\mmix$ are also shown.}
\end{figure}

For $y=0$, this set of solutions reduces to the FP SUSY models with
$m^2 \gg A, M$ discussed
previously~\cite{Feng:1999mn,Feng:1999zg,Feng:2000bp,Feng:2000gh}.
For $x = y= 0$, the set reduces further to mSUGRA/CMSSM with
$A_t(\mgut)=0$, which also exhibits FP behavior for large scalar mass
$m_0^2$.  The mSUGRA/CMSSM case is highlighted in \figref{xy}, along
with two other representative models that will be examined in detail
below.  The blue dashed and green dot-dashed lines correspond to the
special cases where $\mtl^2 = \mtr^2$ at the GUT scale and weak scale,
respectively.

\subsection{Higgs Boson Mass} 

It is well known that radiative corrections to the Higgs boson mass
are enhanced both by large stop masses and significant left-right stop
mixing at the weak scale.  In the limit $\mtl (\mweak) = \mtr
(\mweak)$, the one-loop and dominant two-loop corrections to the Higgs
mass from the stop sector are~\cite{Carena:1995bx}
\begin{equation}
\Delta m_h^2 = \frac{3}{4\pi^2}\frac{m_t^2}{v^2}\left\{ \ln
  \left(\frac{M_S^2}{m_t^2}\right) + \tilde{X}_t + \frac{3}{32\pi^2}
  \frac{m_t^2}{v^2} \left[ 2 \tilde{X}_t \ln \!
  \left(\frac{M_S^2}{m_t^2}\right) + \ln^2 \!
  \left(\frac{M_S^2}{m_t^2}\right) \right] \right\} \, ,
\label{higgsmass}
\end{equation}
where $v \simeq 246~\gev$, $M_S^2$ is given in terms of the physical
stop masses $m_{\tilde{t}_{1,2}}$ by $M_S^2 = (m_{\tilde{t}_1}^2 +
m_{\tilde{t}_2}^2 ) / 2$, and
\begin{equation}
\tilde{X}_t = \frac{\left[A_t (\mweak) - \mu
  \cot\beta\right]^2}{M_S^2} \left[ 1 - \frac{\left[ A_t (\mweak) - \mu
    \cot\beta \right]^2}{12 M_S^2} \right] \ .
\end{equation}
The radiative contribution generated by stop mixing is contained in
$\tilde{X}_t$, which increases with $\at (\mweak)$ up to a maximal
value at $\left[A_t (\mweak) - \mu \cot\beta\right]^2 /M_S^2 =
6$~\cite{Carena:1998wq}.  Although this enhancement of the Higgs mass
depends on the stop mass, it can easily exceed $\sim
10~\gev$~\cite{Heinemeyer:1998np}.

In the focus point scenario $\mtl (\mweak) = \mtr (\mweak)$ does not
generically hold, but it can reasonably be assumed that the radiative
contribution to the Higgs mass from mixing will be increase with
$|\at(\mweak)|$ until $|\at (\mweak)| \sim m_{Q_3,U_3} (\mweak)$.  The
enhancement may be parameterized in terms of the mixing parameter
\begin{equation}
\mmix = \frac{\left|\at (\mweak)\right|}{\sqrt{\frac{1}{2} \left[
    \mtl^2 (\mweak) + \mtr^2 (\mweak) \right]}} =
\frac{\sqrt{y}}{\sqrt{\frac{1}{2} - \frac{3}{2}y}} \ .
\end{equation}
Here we have neglected the contribution from $\mu \cot \beta$,
typically small for natural theories with moderate to large
$\tan\beta$, and used $m_{Q_3,U_3} (\mweak)$ instead of
$m_{\tilde{t}_1,\tilde{t}_2}$ to allow for comparison between models
with varying $m_0$.\footnote{Other definitions of $\mmix$ can be used,
  such as $| \at (\mweak) | /\left[ \frac{1}{2} \left( \mtl + \mtr
    \right) \right]$ or $| \at (\mweak) | /\sqrt{ \mtl \mtr} $. All
  are roughly equivalent for $\mtl (\mweak) \approx \mtr (\mweak)$.}

\Figref{xy} contains contours of $\mmix$ over the range of allowed
value of $x$ and $y$.  Our expression for the radiative corrections to
the Higgs mass are approximately correct along the green (dot-dashed)
line corresponding to $\mtl (\mweak) = \mtr (\mweak)$, and $\mmix =
\sqrt{6}$ is only possible when this condition is approximately
satisfied.  For $\mtl (\mgut) = \mtr (\mgut)$, the maximum value of
$\mmix$ is roughly 1.  As we will see from the numerical analysis of
the next section, even such large, but non-maximal, values of $\mmix$
lead to significant enhancements of the Higgs mass.\footnote{For $\mtl
  (\mweak) \neq \mtr (\mweak)$, the maximal correction to $m_h$ will
  occur for different values of $\mmix$, possibly at smaller values.
  However, in the case of reasonably large $m_0$ that we will consider
  numerically, these effects should appear near the tachyonic stop
  boundary, where the approximate form of $\mmix$ using $m_{Q_3,U_3}
  (\mweak)$ breaks down.}

\section{Numerical Results} 
\label{sec:numerical}

\subsection{Three Representative Cases}
\label{sec:three}

We now determine to what extent the analytic results of the previous
sections are realized when all numerical details are included.  We do
this by choosing three representative models to analyze.

For reference and comparison to previous studies, we include
mSUGRA/CMSSM with $A_t(\mgut) = 0$ as one of these cases.  In
addition, we would like to consider models with significant stop
mixing.  The special case of $\mmix = \sqrt{6}$ is achievable for $y
\approx 0.3$. At this point, both stop masses are roughly 20\% of
their GUT-scale values, which results in a moderate reduction in the
mixing-independent radiative contribution to the Higgs mass.
Moreover, this point is near the tachyonic stop boundary for both
$\mtl^2$ and $\mtr^2$, where the neglected effects from proper
diagonalization of the stop mass matrix, one-loop corrections, and the
RG contributions of the gauginos become important.  Inclusion of these
effects can easily produce tachyonic stop masses in a particular
model, or they may raise one stop mass and thus reduce $\mmix$
significantly.  A detailed analysis of the maximal mixing scenario was
recently made in Ref.~\cite{Brummer:2012ns}.

To avoid these issues, the additional two models we consider are far
from the tachyonic stop boundaries with significant, but non-maximal,
$\mmix$.  Heavy stop masses can still be natural in FP SUSY, and so
even with non-maximal mixing, the radiative contribution to the Higgs
boson mass may be substantial.  The models are:
\begin{eqnarray}
\text{mSUGRA/CMSSM\ with $A_0=0$} &:& \ (x,y) = \left(0, 0 \right) \\ 
\text{Model A} &:& \ (x,y) = \left(\frac{1}{4}, \frac{1}{6} \right) \\ 
\text{Model B} &:& \ (x,y) = \left(\frac{5}{9}, \frac{7}{27} \right) \ ,
\end{eqnarray}
as indicated in \figref{xy}.  Model A has $\mmix = 0.82$ and equal
stop masses at the GUT scale, $\mtl^2 (\mgut) = \mtr^2 (\mgut) =
\frac{3}{4} m_0^2$.  Model B has $\mmix = 1.53$ and equal stop masses
at the weak scale, $\mtl^2 (\mweak) = \mtr^2 (\mweak) = \frac{1}{9}
m_0^2$.  Both models will produce $m_{\tilde{t}_1, \tilde{t}_2} \agt
1~\tev$ for $m_0 \ge 3~\tev$.  We emphasize, however, that models near
the tachyonic stop boundaries will, of course, predict lighter stops,
and are perfectly well-defined possibilities.  We do not consider them
for simplicity, but some of them have near maximal mixing and may
produce the 125 GeV Higgs boson with even less fine-tuning than the
models we consider.

To derive numerical results, we must, of course, specify the complete
SUSY model, including parameters that have little impact on
electroweak symmetry breaking and naturalness.  For concreteness, we
fix all scalar masses (except the stops) to be degenerate with
$\mhu^2$ at the GUT scale, $\tan\beta = 10$, and $\mu > 0$.  Although
$\at^2 (\mgut)$ is fixed, the sign of $\at (\mgut)$ is not.  We set
$\at (\mgut) < 0$; this choice will be discussed further in
\secref{position}. We impose unification of the $A$-terms at the GUT
scale, using $A_0 = \at (\mgut)$.  Both spectra and fine-tuning are
obtained using SOFTSUSY 3.1.7~\cite{Allanach:2001kg}, suitably
modified to include the non-universal boundary conditions and
correlations assumed in these FP models.  The numerical analysis
therefore includes two-loop RGEs, one-loop threshold corrections,
minimization of the electroweak potential at $M_S$, and quadratic
contributions to the Higgs mass at energy scales below the masses of
the superpartners, where they have been integrated out.  There is an
uncertainty in the determination of the Higgs mass, which has been
estimated to be roughly 3 to 5
GeV~\cite{Degrassi:2002fi,Allanach:2004rh} through comparison of
numerical routines.

\begin{figure}[tb]
\includegraphics[width=0.49\columnwidth]{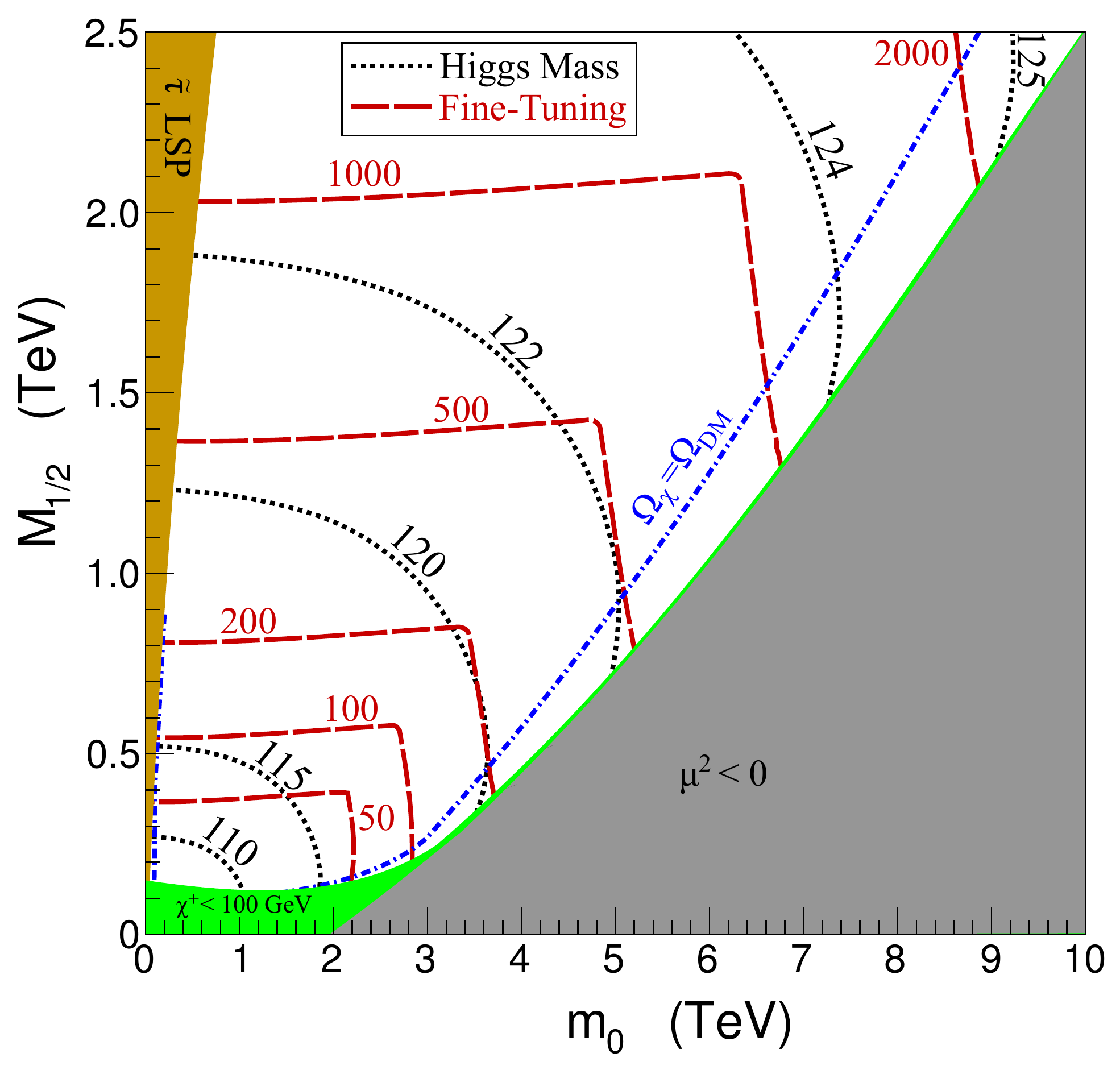}
\hfil
\includegraphics[width=0.49\columnwidth]{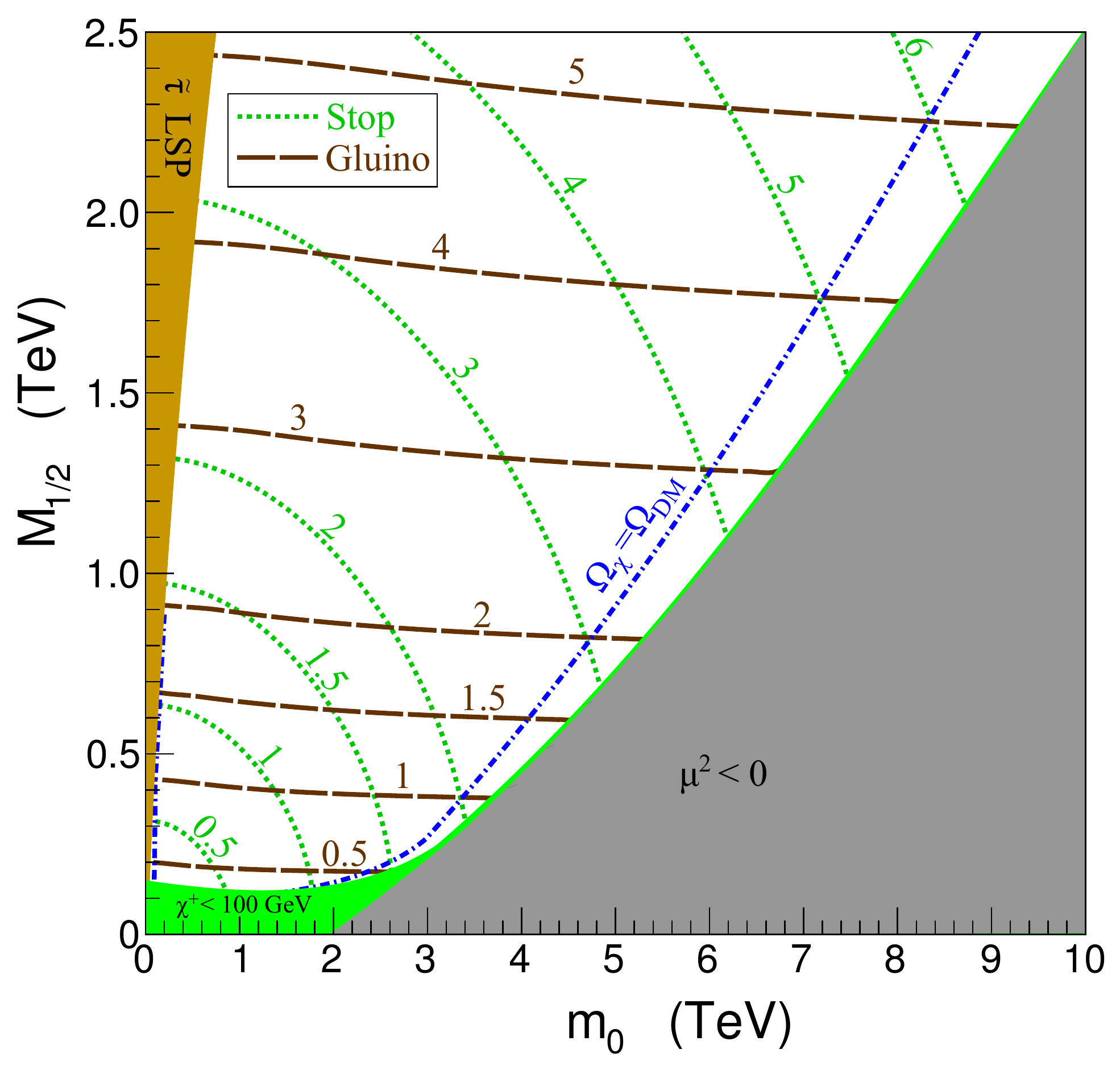}
\vspace*{-.1in}
\caption{\label{fig:cmssm} mSUGRA/CMSSM with $(x,y) = (0,0)$. Left:
  Contours of Higgs boson mass $m_h$ (black dotted) and fine-tuning
  parameter $c$ (red dashed) in the $(m_0, \mgaugino)$ plane.  Right:
  Contours of gluino mass $M_3$ (brown dashed) and lighter stop mass
  $m_{\tilde{t}_1}$ (green dashed) in the $(m_0, \mgaugino)$ plane.
  In both panels, the region where neutralino dark matter has the
  correct thermal relic abundance is given by the blue dot-dashed
  line.  The shaded regions are excluded because electroweak symmetry
  is not broken (gray), charginos are too light (green), or the
  lightest supersymmetric particle is a stau (gold).  For
  definiteness, we assume gaugino mass unification, $\tan\beta = 10$,
  $\mu >0$, and at the GUT scale, the stop masses are defined by the
  FP condition, and all other scalar masses are set to $m_0$.}
\end{figure}

\Figref{cmssm} contains contours of $m_h$ and $c$ (left panel) and
$m_{\tilde{g}}$ and $m_{\tilde{t}_1}$ (right panel) in the $(m_0,
\mgaugino)$ plane for the mSUGRA/CMSSM case with $\at (\mgut) =0$.  In
this case, $c = c_{\mgaugino}$ for low $m_0$, and remains roughly
constant as $m_0$ increases until $c_{m_0}$ becomes the dominant
contribution at large $m_0$, where the contours angle downward.
However, even when fine-tuning due to $m_0$ is dominant, it is greatly
suppressed relative to the na\"{i}ve value of $m_0^2/m_Z^2$.  Taking
the results at face value, we find that it is possible to achieve a
Higgs boson mass of 125 GeV for $m_0 \approx 10~\tev$, $\mgaugino
\approx 2~\tev$, and $c \sim 2000$, an order of magnitude less
fine-tuning than would be required without the FP mechanism. We note,
however, that these conclusions are extremely sensitive to
uncertainties in the experimental measurements and theoretical
calculations of the Higgs mass.  For example, if these effects combine
to imply that we have overestimated the Higgs boson mass by 5 (3)
GeV~\cite{Degrassi:2002fi,Allanach:2004rh}, a 125 GeV Higgs mass
requires $m_0 \approx 3.5\ (5)~\tev$, $\mgaugino \approx 0.5
\ (1)~\tev$, and $c \sim 200\ (500)$.  Within current uncertainties,
then, even the focus point region of mSUGRA/CMSSM with $\at(\mgut) =
0$ may yield the desired Higgs boson mass with fine-tunings not far
below the percent level.

\begin{figure}[tb]
\includegraphics[width=0.49\columnwidth]{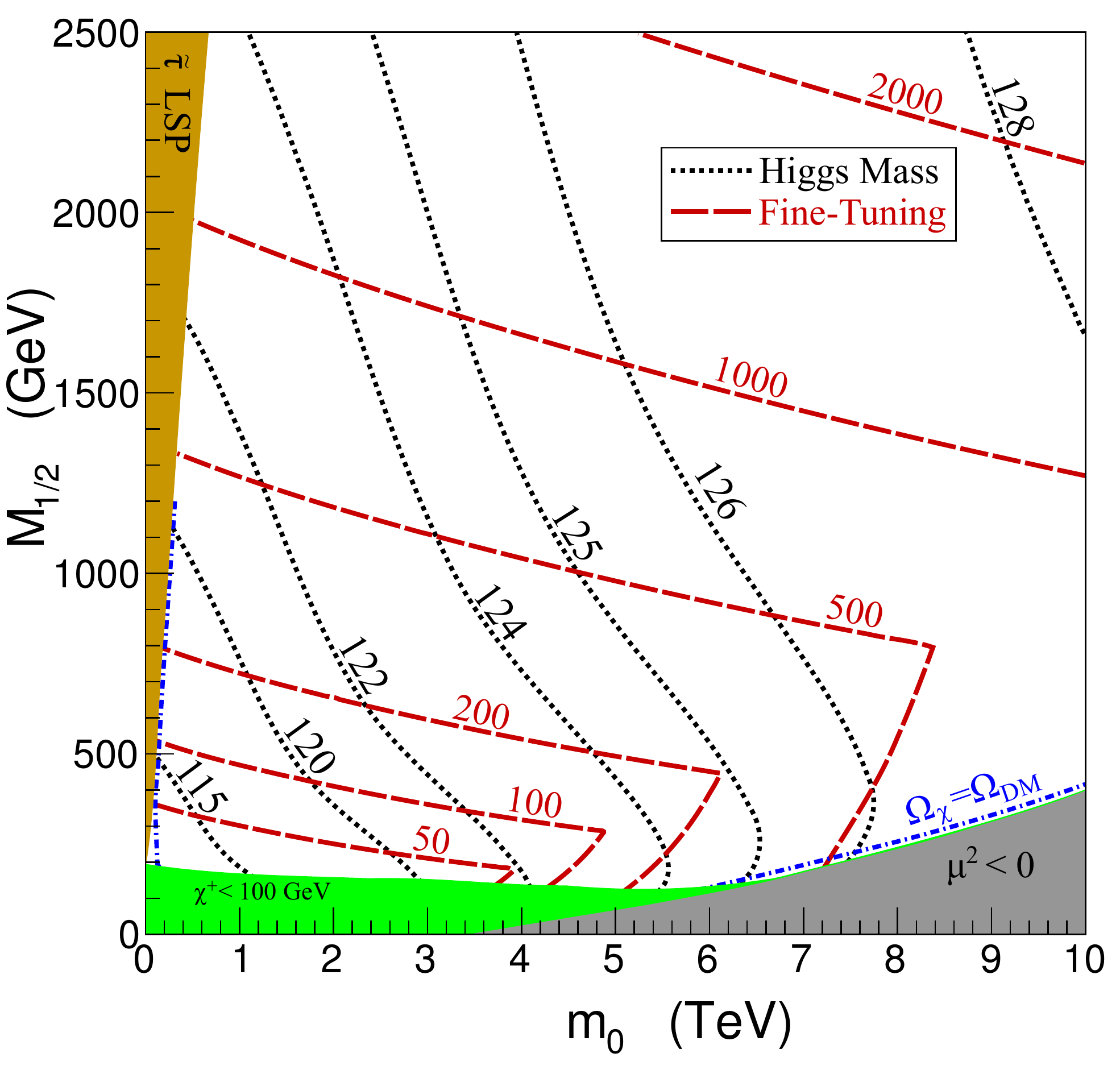}
\hfil
\includegraphics[width=0.49\columnwidth]{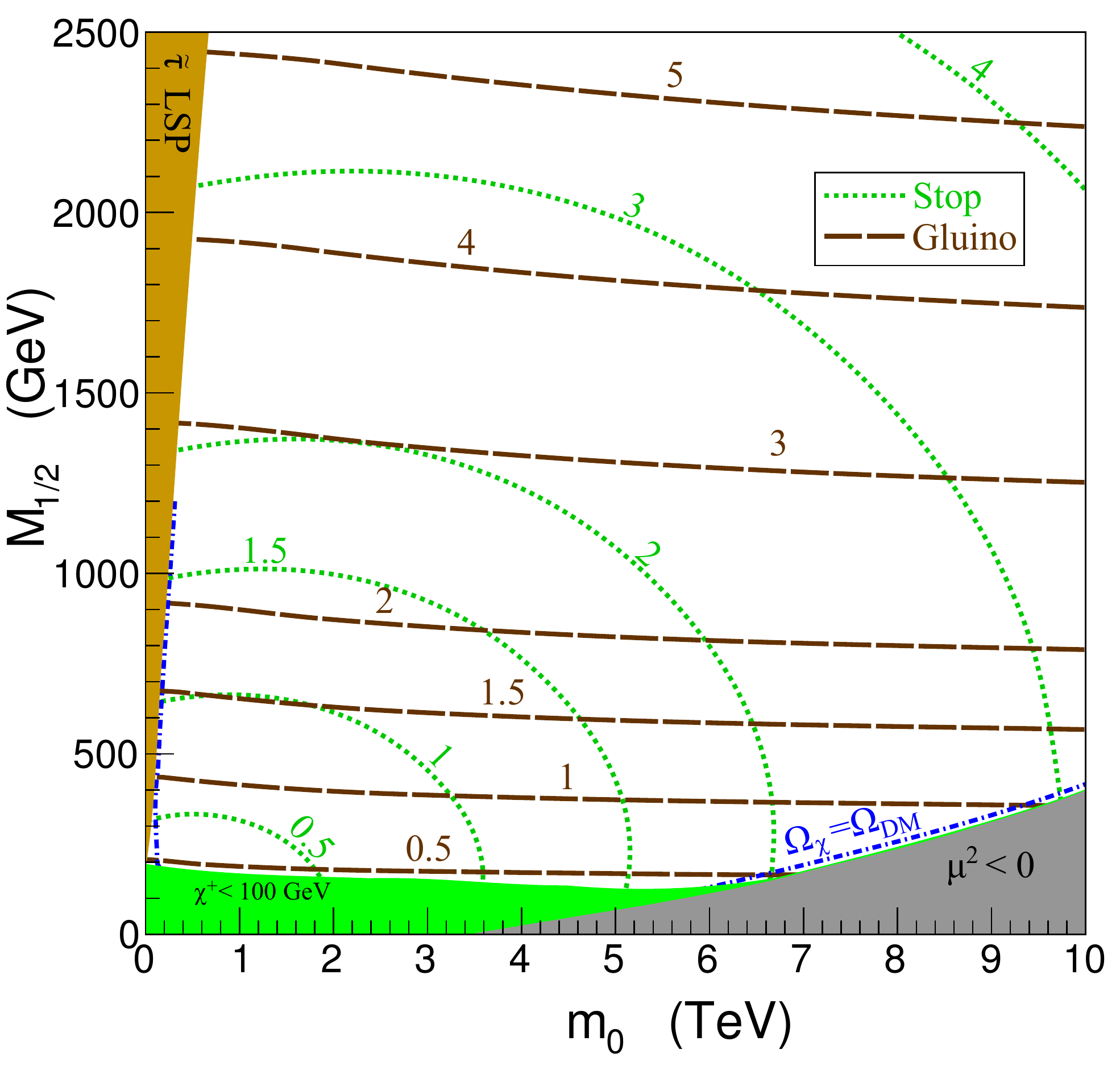}
\vspace*{-.1in}
\caption{\label{fig:unifiedstop} As in \figref{cmssm}, but for Model A
  with $(x,y) = (\frac{1}{4}, \frac{1}{6})$ and GUT-unified stops with
  $\mtl^2 (\mgut) = \mtr^2 (\mgut) = \frac{3}{4} m_0^2$.}
\end{figure}

\begin{figure}[tb]
\includegraphics[width=0.49\columnwidth]{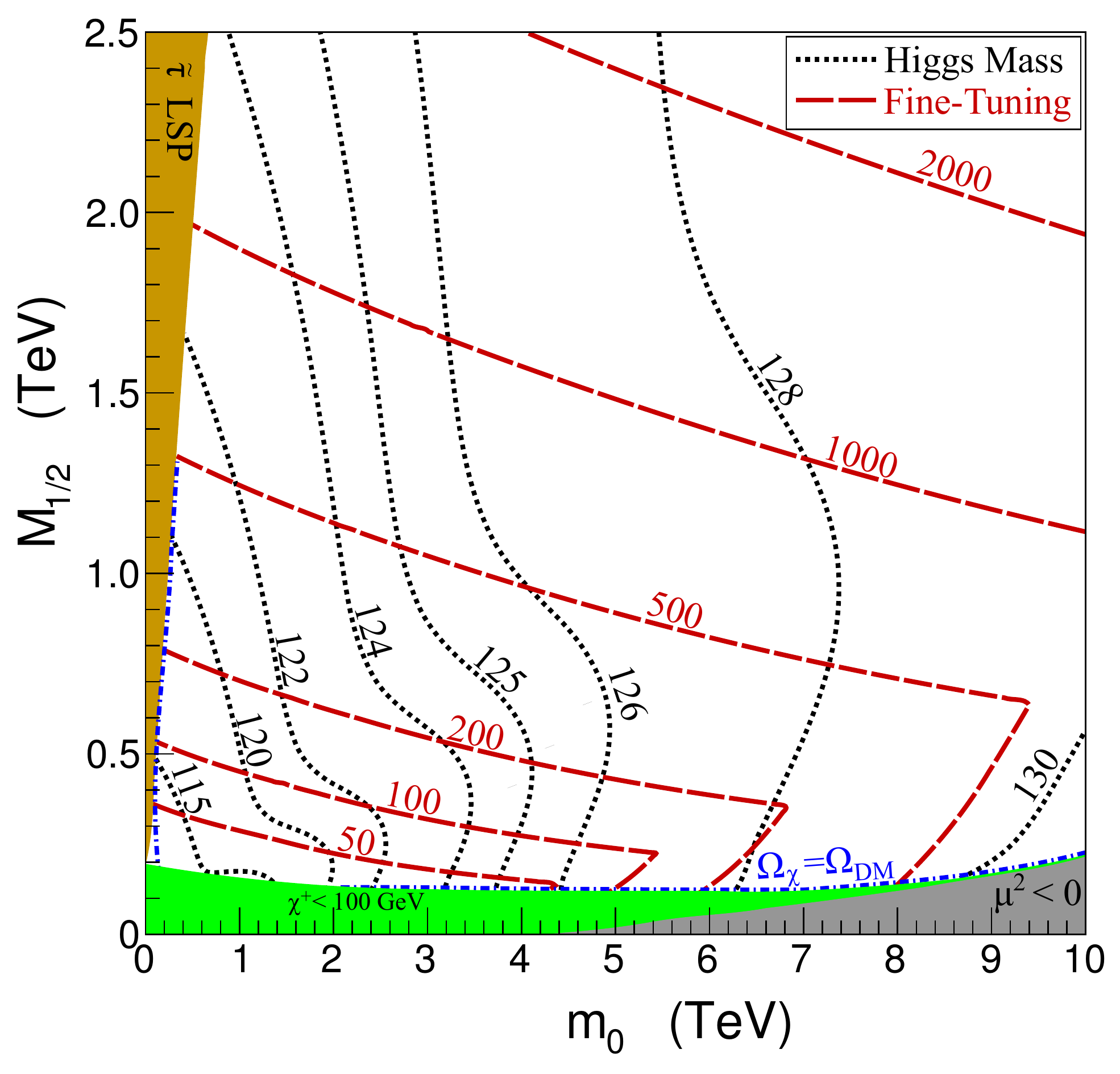}
\hfil
\includegraphics[width=0.49\columnwidth]{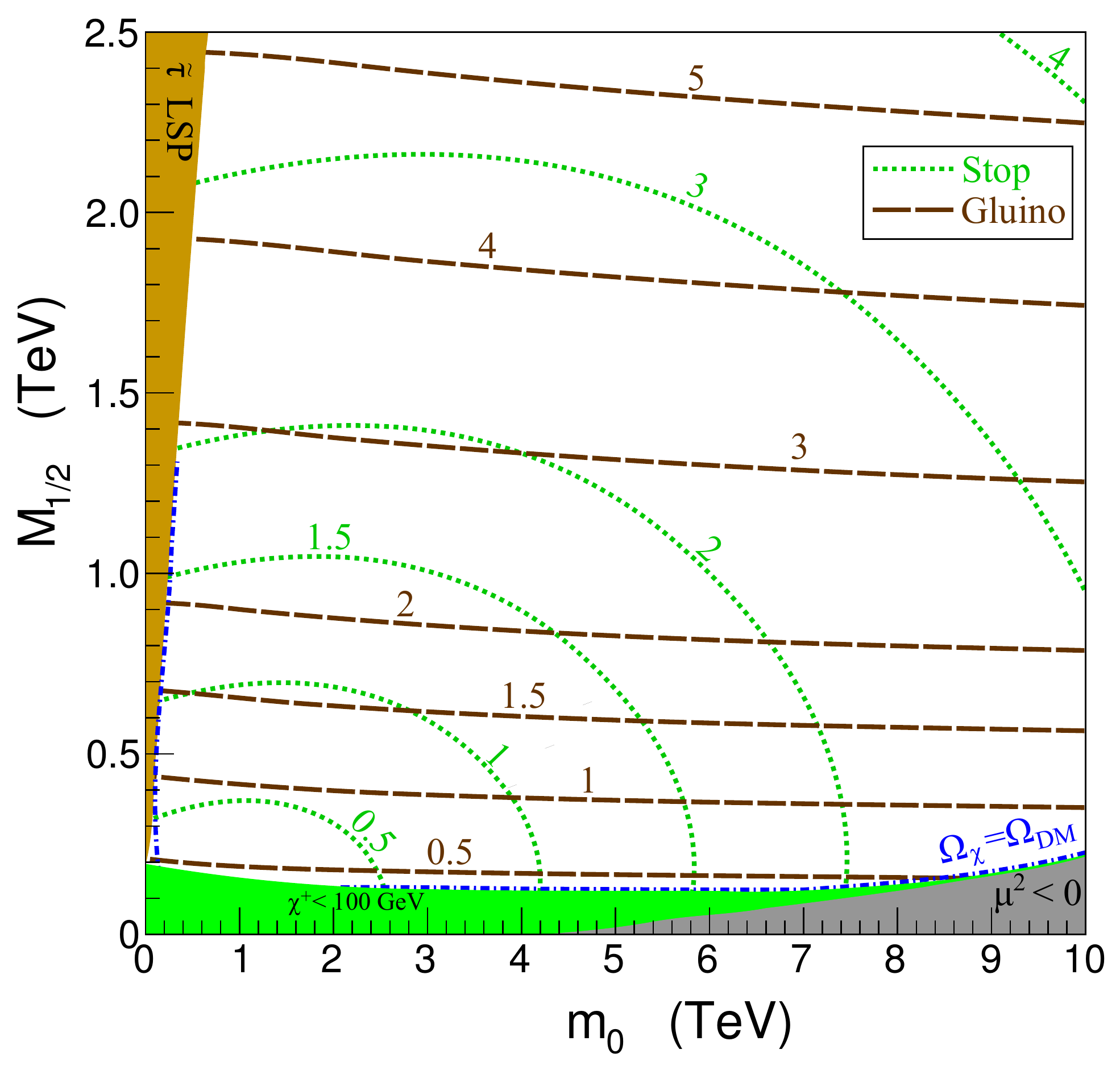}
\vspace*{-.1in}
\caption{\label{fig:largemixing} As in \figref{cmssm}, but for Model B
  with $(x,y) = (\frac{5}{9}, \frac{7}{27})$ and weak-scale-unified
  stops with $\mtl^2 (\mweak) = \mtr^2 (\mweak) = \frac{1}{9} m_0^2$. }
\end{figure}

The results for Models A and B are shown in
\figsref{unifiedstop}{largemixing}, respectively.  The basic behavior
of the fine-tuning contours is similar to the mSUGRA/CMSSM case, but
$c_{\mgaugino}$ is somewhat larger and $c_{m_0}$ slightly smaller at
large $m_0$.  Both effects are larger for Model B.  The more important
effects are the shifts in both the $m_h$ contours and the region with
no viable electroweak minimum.  In both cases, $m_h$ increases
significantly due to contributions from stop mixing, with $m_h =
125~\gev$ found at $m_0 \approx 3-6~\tev$ for Model A and $m_0 \approx
2-4~\tev$ for Model B over the given range of $\mgaugino$.
Furthermore, the position of the $\mu^2 < 0$ region has shifted to
smaller $\mgaugino$ and larger $m_0$, allowing relatively small
gaugino masses at larger values of $m_0$ than would be possible in
mSUGRA/CMSSM.  This effect can best be understood as
\eqref{eq:fullrge} being more approximately true when $\at (\mgut)$ is
increased.

The desire for low fine-tuning motivates consideration of low values
of $\mgaugino$ that yield $m_h = 125~\gev$.  For Model A, the $m_h =
125~\gev$ contour meets the chargino bound at $\mgaugino\approx
200~\gev$, $m_0 \approx 6.5~\tev$, and $c \approx 350$.  The
improvement in fine-tuning over the mSUGRA/CMSSM case is a factor of
5.  Model B demonstrates an even greater improvement, with the
intersection found at $\mgaugino\approx 180~\gev$, $m_0 \approx
3.7~\tev$, and $c < 50$ (fine-tuning of 2\%), a factor of 40
improvement over mSUGRA/CMSSM.  In this region the change in the
gluino mass is marginal between the different models, but the lightest
stop mass is reduced significantly for larger $y$.  In mSUGRA/CMSSM
$m_{\tilde{t}_1} \approx 6~\tev$ when $m_h=125~\gev$, which is reduced
to $m_{\tilde{t}_1} \approx 2~\tev$ in Model A and $m_{\tilde{t}_1}
\approx 900~\gev$ in Model B near the chargino bound.  As in the
mSUGRA/CMSSM case, the fine-tunings are reduced by an order of
magnitude relative to their values without the FP mechanism.  As a
result, percent-level fine-tunings are compatible with $m_h \approx
125~\gev$.

\subsection{Deflection of the Focus Point} 
\label{sec:position}

The formulation of FP scenarios is defined by the condition $\mhu^2
(\mweak) \simeq 0$, even when superpartner masses are significantly
larger.  However, a viable electroweak symmetry breaking minimum
requires $\mhu^2 (\mweak) < 0$.  This behavior could be introduced
directly through an appropriate boundary condition, shifting the FP
boundary condition so that $\mhu^2$ will be small but negative at the
weak scale.  It can also arise dynamically due to sub-dominant terms
in the RG evolution, which generically divert the RG trajectory and
may generate a viable electroweak minimum.

The complete one-loop RGE for the up-type Higgs mass parameter is
\begin{eqnarray}
\nonumber \frac{d \mhu^2}{d \ln Q} &=& \frac{1}{16\pi^2} \Bigg[ 2
  y_t^2 \left( \mhu^2 + \mtl^2 + \mtr^2 + \at^2 \right)  \\
& & - 6 g_2^2 M_2^2 - \frac{6}{5} g_1^2 M_1^2 + \frac{3}{5}
  g_1^2 S \Bigg] \ ,
\label{eq:mhurge}
\end{eqnarray}
where $S$ is given in \eqref{eq:S}.  The FP behavior is governed by
the scalar mass and $\at$ terms of the first line, and the gaugino
masses and $S$ terms in the second line deflect the solution.  The
$M_1$ and $M_2$ terms in the second line of \eqref{eq:mhurge} drive
$\mhu^2$ positive at the weak scale; the contribution of the $S$ term
can be positive or negative, but in mSUGRA/CMSSM the contribution is
positive and further increases $\mhu^2$.  However, additional
contributions are also introduced in the first line of
\eqref{eq:mhurge} by the deflection of $\at^2$ away from the FP
solution.  The RG evolution of $\at$ is given by
\begin{equation}
\frac{d \at}{d \ln Q} = \frac{1}{16\pi^2} 
\left[ 12 y_t^2 \at + 2 y_b^2 \ab + 
\frac{32}{3} g_3^2 M_3 + 6 g_2^2 M_2 + \frac{26}{15} g_1^2 M_1 \right] ,
\label{eq:atrge}
\end{equation}
where the first term is considered in the FP analysis, and the
remaining terms deflect the solution.  This results in a significant
deflection of $\at (\mweak)$ for non-zero gluino mass, producing a
corresponding deflection in $\mhu^2$ that rivals the deflection from
the terms in the second line of \eqref{eq:mhurge}.

In the case of $\at (\mgut) = 0$, RG evolution generates a non-zero
value for $\at (\mweak)$, which in turn drives $\mhu^2$ more negative
at the weak scale.  For scenarios with gaugino mass unification, this
contribution is larger than the direct one-loop contributions from
$M_1$ and $M_2$, producing an overall negative contribution to
$\mhu^2$ at the weak scale.  This must still be balanced against the
positive contribution from $S$ and, for sufficiently large scalar
masses $\mhu^2$, will be positive at the weak scale.  This is the
origin of the phenomenologically-excluded $\mu^2 < 0$ region in
mSUGRA/CMSSM at high $m_0$, which requires larger values of $m_0$ with
increasing $\mgaugino$, as shown in \figref{cmssm}.

For $\at (\mgut) \agt M_3$, the contribution to $\mhu^2$ depends upon
the relative sign of $\at (\mgut)$ and $M_3$.  The deflection of
$\at^2 (\mweak)$ is $\Delta \at^2 (\mweak) \sim - \at (\mgut) M_3$,
with a negative deflection of $\mhu^2$ when $\Delta \at^2 (\mweak) >
0$.  When $\at (\mgut)$ and $M_3$ have the same sign this results in
no viable electroweak minimum, but when the signs are opposite, a
viable minimum is reached similarly to the $\at (\mgut) = 0$ case.
This is the origin of the condition $\at (\mgut) < 0$ in
\secref{three}.  For $\at (\mgut) \neq 0$, the $\mu^2 < 0$ region also
shifts to higher values of $m_0$. This is because the RG contributions
from direct deflections of $\mhu^2$ are unchanged, but $\Delta \at^2
(\mweak)$ is increased relative to the $\at (\mgut)=0$ case.

\section{Implications for Colliders and Dark Matter}
\label{sec:experimental}

The most immediate experimental implication of the FP SUSY models
discussed here is that they naturally predict a SM-like Higgs boson in
the currently allowed range around $m_h \approx 125~\gev$.  If these
models are realized in nature, the Higgs boson should be discovered in
the very near future with properties consistent with those of a
SM-like Higgs boson.

The FP SUSY models discussed here also naturally explain the
non-observation of superpartners at the LHC so far.  In FP models, the
squarks and sleptons of the first two generations barely RG evolve,
and so have physical masses essentially set by their value at the GUT
scale. As noted above, electroweak symmetry breaking is highly
insensitive to these masses, and so they are not constrained by the FP
mechanism.  However, under the assumption that they are $\sim m_0$, in
all the scenarios we consider, for $m_h \approx 125~\gev$, the squarks
of the first two generations are in the multi-TeV region, well beyond
the current bound of $m_{\tilde{q}} \agt
1.4~\tev$~\cite{ATLAS-CONF-2012-033,CMS-PAS-SUS-12-002}.  The stop
masses do RG evolve in these models, and they may be somewhat lighter,
but they are also beyond current bounds, which are much weaker for
third generation squarks~\cite{Aad:2011cw,Aad:2012unmarked}.

In the future, probably the most promising avenues for collider
discovery are stop and gluino searches.  As noted in
\secref{numerical}, the most natural FP scenario we have considered in
detail (Model B) has large stop mixing and relatively light stops in
the range $m_{\tilde{t}} \sim 1~\tev$.  As emphasized in
\secref{three}, however, for simplicity, we have purposely avoided
models near the tachyonic stop boundaries, where the stop mixing is
even higher.  These will produce models with lighter stops, and quite
possibly even less fine-tuning.  Such stops will be within reach of
future LHC analyses.  Of course, light stops are a general feature of
many natural SUSY theories.  The new feature of FP models with light
stops is that the fine-tuning is significantly reduced, even including
the fine-tuning with respect to the stop mixing parameter, and the
particle content is minimal, and so conventional MSSM searches are
applicable.  Future gluino searches are also promising.  As discussed
in \secref{focuspoints}, for reasons related to the general structure
of the RGEs, it is quite natural for the scalars to participate in the
FP mechanism, but not the gauginos.  For this reason, requiring less
fine-tuning than 1 part in 1000 implies $m_{\tilde{g}}\alt 3 -
4~\tev$.  The relatively light stops also imply that gluinos, if
produced, will decay dominantly through top- and bottom-rich cascade
decays through off-shell (or even on-shell)
stops~\cite{Chattopadhyay:2000qa,Kadala:2008uy,Aad:gluino,CMS-PAS-SUS-11-028}.

The phenomenology of dark matter for the large $\at (\mgut)$ models we
consider is more complicated than in mSUGRA/CMSSM FP scenarios.  For
the mSUGRA/CMSSM case, there is an excellent thermal relic dark matter
candidate, a Bino-like neutralino with a significant Higgsino
component.  It has the correct thermal relic density in the region of
parameter space with $m_h \approx 125~\gev$, and will either be
detected or excluded by direct detection searches in the near
future~\cite{Feng:2011aa}.  For large $\at (\mgut)$, the preferred
region with $m_h \approx 125~\gev$ is at lower $m_0$, but, as
discussed in \secref{position}, the $\mu^2 < 0$ region is found at
much larger $m_0$. The result is that for regions of parameter space
with $m_h \approx 125~\gev$, $\mu$ is significantly above $M_1$, the
dark matter is nearly pure Bino, and its thermal relic density is
typically too large.  This may be fixed when $m_\chi \approx m_h / 2$
and the annihilation is enhanced by the (SM-like) Higgs funnel.
Unfortunately, this slice of parameter space is located at $\mgaugino
\sim 150 - 200~\gev$, which is inconsistent with the recent ATLAS
gluino mass bound of $m_{\tilde{g}} \agt
900~\gev$~\cite{ATLAS-CONF-2012-033} for decoupled squarks.  The bound
requires $\mgaugino \agt 400~\gev$, and at such high gaugino masses,
the Higgs resonance is not in effect, and relic neutralinos are
overabundant.

There are, however, several possible solutions to this dark matter
problem.  First, one may, of course, always invoke $R$-parity
violation or allow the neutralino to decay to another, lighter
supersymmetric particle, such as a
gravitino~\cite{Pagels:1981ke,Feng:2003xh} or
axino~\cite{Rajagopal:1990yx,Covi:1999ty}. In the standard formulation
of gravity mediation, the gravitino mass is typically of the order of
the masses of SM superpartners, but light gravitinos can be introduced
through a non-standard K\"{a}hler potential~\cite{Kersten:2009qk}.

More satisfying, however, is the observation that all of the analysis
of the previous paragraph relies heavily on the assumption of gaugino
mass unification.  The gaugino masses play a sub-leading role in the
FP analysis, and non-unified gaugino masses are perfectly possible in
FP models.  A thermal relic may be restored either by reducing $M_1$
to produce a light neutralino, which increases its annihilation cross
section, or increasing $M_1$ to values closer to $\mu$, thereby
increasing the Higgsino component of the neutralino, and with it, the
annihilation cross section.  In either case, the scattering cross
section can be expected on general grounds to rise with the
annihilation cross section. Although more work is required, the
prospects for both direct and indirect dark matter searches for
neutralino dark matter should be excellent; for example, in the
Bino-Higgsino case, they should be similar to those for mSUGRA/CMSSM
FP
neutralinos~\cite{Baer:2008uu,Feng:2010ef,Gogoladze:2010ch,Akula:2011dd}.
Irrespective of shifts in $M_1$, considering a point with $m_h =
125~\gev$ and $m_{\tilde{g}} = 900~\gev$ produces a fine-tuning of $c
\approx 300$ for Model A and $c \approx 110$ for Model B, and these FP
scenarios provide a natural possibility for realizing $m_h \approx
125~\gev$ in the MSSM consistent with WIMP thermal relic dark matter.

\section{Conclusions} 
\label{sec:conclusions}

Many of the most cherished virtues of weak-scale SUSY, such as gauge
coupling unification, radiative electroweak symmetry breaking, and the
heavy top quark, are dynamically generated. In this study, we have
examined FP SUSY, in which naturalness is also dynamically generated.
This framework is motivated by the desire to find natural
supersymmetric theories with 125 GeV Higgs masses in the MSSM, without
extending the theory with additional field content or invoking
low-scale SUSY-breaking mediation.

We have extended earlier works on FP SUSY to analyze the possibility
that both scalar masses and $A$-terms are multi-TeV and hierarchically
larger than all other soft parameters.  We find that the FP mechanism
may be realized if the soft parameters are in the relation $\mhu^2 :
\mtr^2 : \mtl^2 : \at^2 = 1 : 1+x - 3y : 1-x : 9y $, where $x$ and $y$
are in the ranges given in \eqref{eq:range}. The FP mechanism is
independent of all other scalar masses, and of the gaugino masses,
provided they are smaller than these.

We examined three particular choices for $(x,y)$ in detail.  The
results may be very roughly summarized, and contrasted with previously
results, as follows.  For general models with large stop masses, there
are both quadratic contributions $(6/8\pi^2) y_t^2 m_{\tilde{t}}^2$
and large logarithm-enhanced contributions $\sim (6/8\pi^2) y_t^2
m_{\tilde{t}}^2 \ln (\mgut/m_{\tilde{t}})$.  Without the FP mechanism,
the large log terms dominate.  A 125 GeV Higgs mass may be achieved
with 10 TeV stop masses and no mixing, leading to a fine-tuning of
roughly 1 part in $10^4$, or with highly mixed, few TeV stops, with a
fine-tuning of roughly 1 part in 1000~\cite{Hall:2011aa}.  The FP
mechanism suppresses the leading log terms, and so reduces fine-tuning
by roughly an order of magnitude.  Previously, FP models with no stop
mixing achieved a 125 GeV Higgs with 10 TeV stops and fine-tuning of
around 1 part in 1000~\cite{Feng:2011aa}.  In this work, we have found
new FP models with significant stop mixing, where the 125 GeV Higgs
bosons are achieved with fine-tuning of 1 part in 100.  

To our knowledge, the models presented here are the first with minimal
field content and SUSY-breaking mediated at the GUT scale that
accommodate a 125 GeV Higgs boson with only percent-level fine-tuning.
General implications for SUSY searches at colliders and dark matter
experiments have been summarized in \secref{experimental}.  The model
framework provides a simple extension of the mSUGRA/CMSSM boundary
conditions that simultaneously preserves naturalness, accommodates the
new Higgs boson constraints, and predicts superpartners within reach
of the LHC, and is therefore an ideal framework for presenting new
results from LHC searches.

The motivation to find natural and simple theories consistent with a
125 GeV Higgs boson has led us to a predictive relation between soft
parameters in the top sector.  If the predictions of FP models
continue to be born out, it will become increasingly interesting to
explore what UV frameworks may naturally yield the required relations.
Such a study is beyond the scope of this work, but we close with some
speculations.  The general solution presented here allows for the
possibility for FP behavior for non-zero $\at (\mgut)$, provided there
is some splitting between the masses $m_{H_u}^2$, $\mtl^2$, and
$\mtr^2$ at the GUT scale.  Such splitting is generically possible in
SUSY theories derived from superstring models with SUSY-breaking
arising from the dilaton/moduli sector of the theory, and indeed there
is a direct connection between the size of the $A$-term and the mass
splitting between scalars in such
frameworks~\cite{Ibanez:1992hc,Brignole:1997dp}.  It is interesting to
note that the required boundary conditions could arise either purely
from the dilaton/moduli sector of such a theory, or in combination
with a direct $F$-term mediation scheme, providing a UV explanation
for this class of models.

\section*{Acknowledgments} 

We thank Konstantin Matchev, Michael Ratz, and Yuri Shirman for
enlightening discussions.  JLF and DS are supported in part by NSF
grant PHY-0970173.

\bibliography{bibafocus1}{}

\end{document}
